# *In-situ* strain tuning of the Dirac surface states in Bi$_2$Se$_3$ films

David Flötotto[1,2]*, Yang Bai[1,2], Yang-Hao Chan[3], Peng Chen[1,2,4], Xiaoxiong Wang[5], Paul Rossi[6], Cai-Zhi Xu[1,2], Can Zhang[1,2], Joe A. Hlevyack[1,2], Jonathan D. Denlinger[4], Hawoong Hong[7], Mei-Yin Chou[3,8,9], Eric J. Mittemeijer[6], James N. Eckstein[1,2] and Tai-Chang Chiang[1,2,9]*

[1]Department of Physics, University of Illinois at Urbana-Champaign, Urbana, IL 61801, USA

[2]Frederick Seitz Materials Research Laboratory, University of Illinois at Urbana-Champaign, Urbana, IL 61801, USA

[3]Institute of Atomic and Molecular Sciences, Academia Sinica, Taipei 10617, Taiwan

[4]Advanced Light Source, Lawrence Berkeley National Laboratory, Berkeley, CA 94720, USA

[5]College of Science, Nanjing University of Science and Technology, Nanjing 210094, China

[6]Max Planck Institute for Intelligent Systems, Heisenbergstraße 3, D-70569 Stuttgart, Germany

[7]Advanced Photon Source, Argonne National Laboratory, Argonne, IL 60439, USA

[8]School of Physics, Georgia Institute of Technology, Atlanta, GA 30332, USA

[9]Department of Physics, National Taiwan University, Taipei 10617, Taiwan

*Correspondence to: D. Flötotto (flototto@illinois.edu) and T.-C. Chiang (tcchiang@illinois.edu; Tel. (217) 333-2593)

**Abstract**

Elastic strain has the potential for a controlled manipulation of the band gap and spin-polarized Dirac states of topological materials, which can lead to pseudo-magnetic-field effects, helical flat bands and topological phase transitions. However, practical realization of these exotic phenomena is challenging and yet to be achieved. Here, we show that the Dirac surface states of the topological insulator $Bi_2Se_3$ can be reversibly tuned by an externally applied elastic strain. Performing *in-situ* x-ray diffraction and *in-situ* angle-resolved photoemission spectroscopy measurements during tensile testing of epitaxial $Bi_2Se_3$ films bonded onto a flexible substrate, we demonstrate elastic strains of up to 2.1% and quantify the resulting reversible changes in the topological surface state. Our study establishes the functional relationship between the lattice and electronic structures of $Bi_2Se_3$ and, more generally, demonstrates a new route toward momentum-resolved mapping of strain-induced band structure changes.

Because of the inherent coupling of the electronic structure with the crystal lattice, the application of strain offers a promising pathway to modify the electronic structure of materials. As such, the application of an elastic strain not only has the potential for providing an efficient reversible engineering control of the functional properties of materials,[1-5] but also permits access to novel emergent phases[6] as well as exotic physical phenomena arising from strain-induced lattice-symmetry breaking.[7-9] For topological insulators (TI)[10] specifically, theoretical studies indicate that the bulk band gap and the spin-polarized Dirac surface states can be tailored by elastic strain,[11-16] thus inspiring hopes to use elastic strain as an *in-situ* equivalent of chemical-composition tuning for the implementation of "straintronic" devices.[14] However, the realization of these phenomena is advantageous both from the fundamental and applications perspectives. The challenges in achieving strain control lie in the difficulty of controllably applying strain, characterizing the type and magnitude of the induced strain, and simultaneously measuring the resulting changes of the surface electronic band structure. Intrinsic stresses/strains induced by the lattice mismatch to a substrate into typical TI thin film samples relax within a few layers of film growth due to the weak inter layer bonding,[17] and only low or local strains have been realized.[18-20] Due to these limitations, recent experiments have only probed the (local) density of states in the vicinity of dislocations patterns by scanning tunneling microscopy[18] or strained polycrystalline films through transport measurements[20], but here it is nearly impossible to experimentally disentangle the effect of strain on the topological surface state from other complicating factors such as the presence of electric fields at grain boundaries[18], morphological changes[21] and bulk state[20]. Consequently, the direct experimental quantification of the effects of strain on the topological surface state remains elusive.

To advance the application of topological materials in the field of "straintronics", we have developed an approach that enables angle-resolved photoemission spectroscopy (ARPES) and x-ray diffraction (XRD) measurements during tensile testing of ultrathin *epitaxial* $Bi_2Se_3$ films bonded onto conductive polyimide foils (Kapton®). This setup not only allows systematic *in-situ* control of the strain state at a macroscopic length scale, but also enables a direct, momentum-resolved quantification of the *reversible* strain-induced energy shift of the topological Dirac point, which is simply impossible with traditional approaches based on mismatched epitaxial film growth.

Films of 10- and 12-QL $Bi_2Se_3(0001)$ are epitaxially grown onto thoroughly cleaned $Al_2O_3(0001)$ substrates and capped with a 60 nm thick Nb layer. Each sample (~3x3 mm$^2$) was flipped over and glued with the Nb layer downward onto 50- or 125-μm thick conductive Kapton foils, as extensively used for the fabrication of flexible electronic devices (*22*). The $Al_2O_3/Bi_2Se_3/Nb$/Kapton samples are then mounted on a specially designed strain sample stage for XRD and ARPES measurements to determine the lattice deformation and the band structure of $Bi_2Se_3$ as a function of the imposed elongation of the Kapton foil. Finally, the $Al_2O_3$(0001) substrate, providing protection against contamination of the $Bi_2Se_3$ film, is mechanically cleaved away *in situ* to expose a fresh $Bi_2Se_3$ surface. The cleavage occurs exclusively at the $Bi_2Se_3/Al_2O_3$ interface, leaving the $Bi_2Se_3$ film with a thickness predetermined by the growth process on the Kapton foil. The experimental setup for accommodating this special "flip-chip" sample assembly and a schematic of the crystallographic orientations of the epitaxial $Bi_2Se_3$ films are shown in Fig. 1.

The strain state of the ultrathin $Bi_2Se_3$ film was determined by recording reciprocal-space x-ray maps for increasing tensile strain imposed on the Kapton foil along $x$, $\varepsilon_{Kapton}$. Examples of the detector images (Fig. 2a) show that, with increasing $\varepsilon_{Kapton}$, the 015 reflection of $Bi_2Se_3$ shifts to smaller $Q_x$ and larger $Q_z$, as a consequence of the imposed expansion of the $Bi_2Se_3$ lattice along $x$ and the accompanied Poisson contraction along $z$, respectively. The strain components $\varepsilon_x$ and $\varepsilon_z$ as a function of $\varepsilon_{Kapton}$ are summarized in Fig. 2b for experiments performed under different environments and at different temperatures. Under all conditions, both $\varepsilon_x$ and $\varepsilon_z$ initially change linearly with $\varepsilon_{Kapton}$ and reach maximum values of 2.1% and -0.9%, respectively, for the test conducted at 90 K. The linear behavior up to $\varepsilon_{Kapton}$ ~3% suggests an elastic deformation of the $Bi_2Se_3$ crystal lattice. Indeed, when the strain applied to the Kapton is released, the reflections essentially shift back to their initial positions (Fig. 2c and d), evidencing that enormous macroscopic *elastic* deformations can be realized in ultrathin $Bi_2Se_3$ films. For $\varepsilon_{Kapton}$ exceeding 3%, the strain in the $Bi_2Se_3$ film does not increase any further, and the reflections broaden significantly, implying the onset of plastic deformation. For even larger $\varepsilon_{Kapton}$ of ~4%, the strain in the $Bi_2Se_3$ film starts to relax due to the development of cracks as visible in the optical micrographs (inset in Fig. 2b). The slightly larger extent of the linear elastic regime and the enhanced compressive strain along $z$, $\varepsilon_z$, for the test conducted at low temperatures can be attributed to enhanced van der Waals bonding along the *c*-axis at low temperatures, as reported for the related material $Bi_2Te_3$.[23]

Figure 3a and b show ARPES and corresponding second-derivative maps along $\bar{\Gamma}$-$\bar{K}$ of a $Bi_2Se_3$ film at various $\varepsilon_{Kapton}$. The map at $\varepsilon_{Kapton}$ = 0% shows the topological surface states with the Dirac point at ~0.34 eV, which is typically observed for $Bi_2Se_3$ films exhibiting electron doping due to Se vacancies.[24] With increasing $\varepsilon_{Kapton}$ and the accompanied elastic deformation of the $Bi_2Se_3$ lattice, the position of the Dirac point moves to smaller energies (Fig. 3a, b). These shifts are also evident from the energy distribution curves at $\bar{\Gamma}$ (Fig. 3c). To quantify the observed shifts, the dispersion relations for the topological surface states are extracted by curve fitting[24] (see Supporting Information) as shown by the dashed curves overlaying the data in Fig. 3a. For $\varepsilon_{Kapton} \leq 3\%$, the Dirac point shifts approximately linearly with $\varepsilon_{Kapton}$ to smaller energies by up to -35 meV (Fig. 3d). As with the XRD measurements, the shifts of the Dirac point are reversible within this range of strain as demonstrated by the data presented in Fig. 3d and e, where the Dirac point shifts to smaller energies upon straining but returns to the original position upon unstraining. For $\varepsilon_{Kapton} > 3\%$, the shift of the Dirac point becomes nonlinear (Fig. 3d), in agreement with the relaxation of the film strain revealed by our XRD measurements (cf. Fig 2b).

The experimental results can be directly compared with first-principles calculations for $Bi_2Se_3$ films subjected to increasing $\varepsilon_x$, where the strain components $\varepsilon_y$ and $\varepsilon_z$ are given by the Poisson contraction of Kapton and by energy minimization, respectively. These conditions reflect fairly accurately the strain state of the $Bi_2Se_3$ film during tensile testing, as confirmed by the good agreement of the resulting $\varepsilon_z/\varepsilon_x$ ratio with that determined by XRD at 90 K (-0.41 vs. -0.43; cf. Fig. 2b). Figure 4a shows the calculated band structure for a 6-QL $Bi_2Se_3$ film relative to the Dirac point as a function of $\varepsilon_x$. Strain substantially affects the width and energy positions of the valence and conduction bands. For increasing $\varepsilon_x$, the top of the valence band at $\bar{\Gamma}$ remains essentially stationary, which can be attributed to its strong connection to the Dirac point by analytic continuation of the electronic wave function over a small energy difference. Simultaneously, the increasing compressive strain along $z$ with increasing $\varepsilon_x$ enhances the inter-quintuple-layer coupling of the Se $p_z$ bands, which causes the vertical band gap at $\bar{\Gamma}$ to increase and the conduction band minimum dominated by the antibonding $p_z$ bands of Se to shift upwards relative to the Dirac point.[11-13, 15, 16] However, to provide a direct comparison to the electron doped films in our experiment with a somewhat populated conduction band (see Supporting Information), the theoretical results are re-plotted with the conduction band minimum (or the Fermi level of the doped sample) as the new reference point (Fig. 4b). It then follows that the Dirac point shifts to smaller energies with increasing $\varepsilon_x$. Our combined XRD and ARPES results directly demonstrate the impact of strain on the position of the Dirac point, and moreover, allow us to quantify the magnitude of the shift to be -37 meV per -1% strain along z, comparable to the theoretical value of -29 meV at equivalent lattice deformations (Fig 4c).

Our demonstrated ability to follow the evolution of the surface electronic structure of a thin film as a function of strain by direct ARPES measurements during tensile testing is crucial for advancing materials design, modification, and control. The potential is enormous as evidenced by the impact of high-pressure research that has opened an entire new field in materials science.[25] Our novel approach using a Kapton foil as the strain transmitting medium in combination with a special flip-chip technique for preparing freshly cleaved surfaces enables us to show that the topological surface states of $Bi_2Se_3$ can be reversibly tuned over an elastic strain range up to $\varepsilon_x$ ~2.1%, a level which is simply impossible to achieve in bulk samples due to their much lower yield strength as compared to that of their spatially confined thin film complements. The approach not only has important implications for the application of topological materials in the field of thin-film straintronics, but also opens new routes to systematically explore the role of strain-induced band structure changes for a wide range of fundamental physical phenomena such as superconductivity[7,9], charge density wave formation[26] and direct/indirect band gap transitions[2].

## ASSOCIATED CONTENT

Supporting Information: Additional information about the sample preparation, the experimental setup, the DFT calculations, the fitting of the topological surface state as well as its stability over time.

## AUTHOR INFORMATION

### Corresponding Authors

*E-mail address: D. Flötotto (flototto@illinois.edu) and T.-C. Chiang (tcchiang@illinois.edu)

### Author Contributions

D.F. and T.-C.C. organized the project. D.F., P.C. C.X., and J.D. performed the ARPES measurements. D.F., Y.B., C-Z.X., J.A.H and J.N.E. designed and prepared the samples. D.F., P.R., Y.B., E.J.M. and H.H. performed the XRD measurements. D.F. and T.-C.C. designed the strain sample holder. X.W., Y.-H. C. and M.-Y.C. performed the first-principles calculations. All authors have participated in discussions of the results and contributed to the manuscript. All authors have given approval to the final version of the manuscript.

### Competing interests

The authors declare no competing interests.

### Acknowledgments

This research is supported by (*i*) the US National Science Foundation (Grant No. DMR-17-09945 for T.-C.C. and No. DMR-11-05998 for J.N.E.), (*ii*) the Deutsche Forschungsgemeinschaft (FL 974/1-1), (*iii*) the U.S. Department of Energy, Office of Science,

Office of Basic Energy Sciences, Division of Materials Science and Engineering (Award No. DE-AC02-07CH11358 for J.N.E.) (*iv*) the Thematic Project at Academia Sinica, (*v*) the National Natural Science Foundation of China (No. 11204133), and (*vi*) the Fundamental Research Funds for the Central Universities of China (No. 30917011338). The Advanced Light Source is supported by the Director, Office of Science, Office of Basic Energy Sciences, of the US Department of Energy under Contract No. DE-AC02-05CH11231. Argonne National Laboratory and work performed at the Advanced Photon Source at Argonne National Laboratory was supported by the U.S. Department of Energy (DOE), Basic Energy Sciences, under Contract No. DE-AC02-06CH11357. Part of the work was carried out in the Central Research Facilities of the Frederick Seitz Materials Research Laboratory.

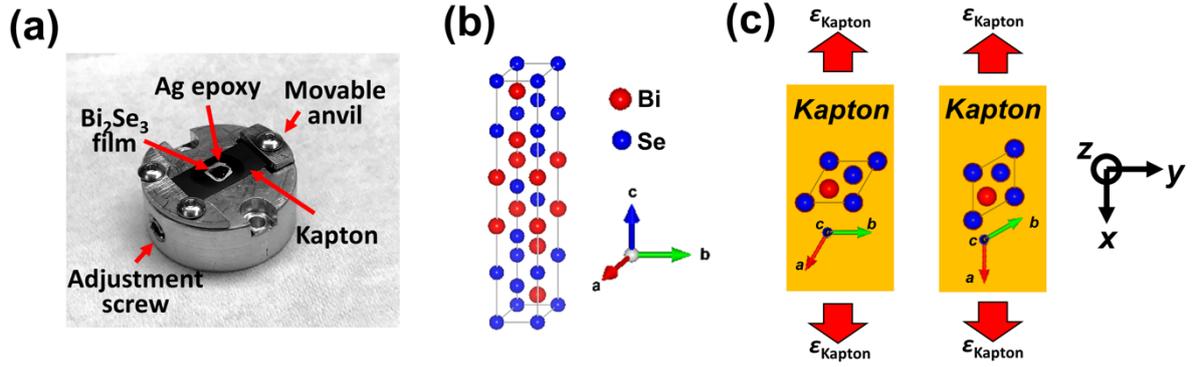

**Figure 1.** (a) Photo of the strain sample holder. The strain is applied by turning the adjustment screw with a wobble stick, thus moving the anvil and thereby straining the conductive Kapton foil and the attached epitaxial $Bi_2Se_3$ film. (b) Unit cell of $Bi_2Se_3$. (c) Schematic drawings of the $Bi_2Se_3$/Kapton samples showing the two choices of orientations of the $Bi_2Se_3$ films with respect to the direction of elongation of the Kapton foil (indicated by red arrows).

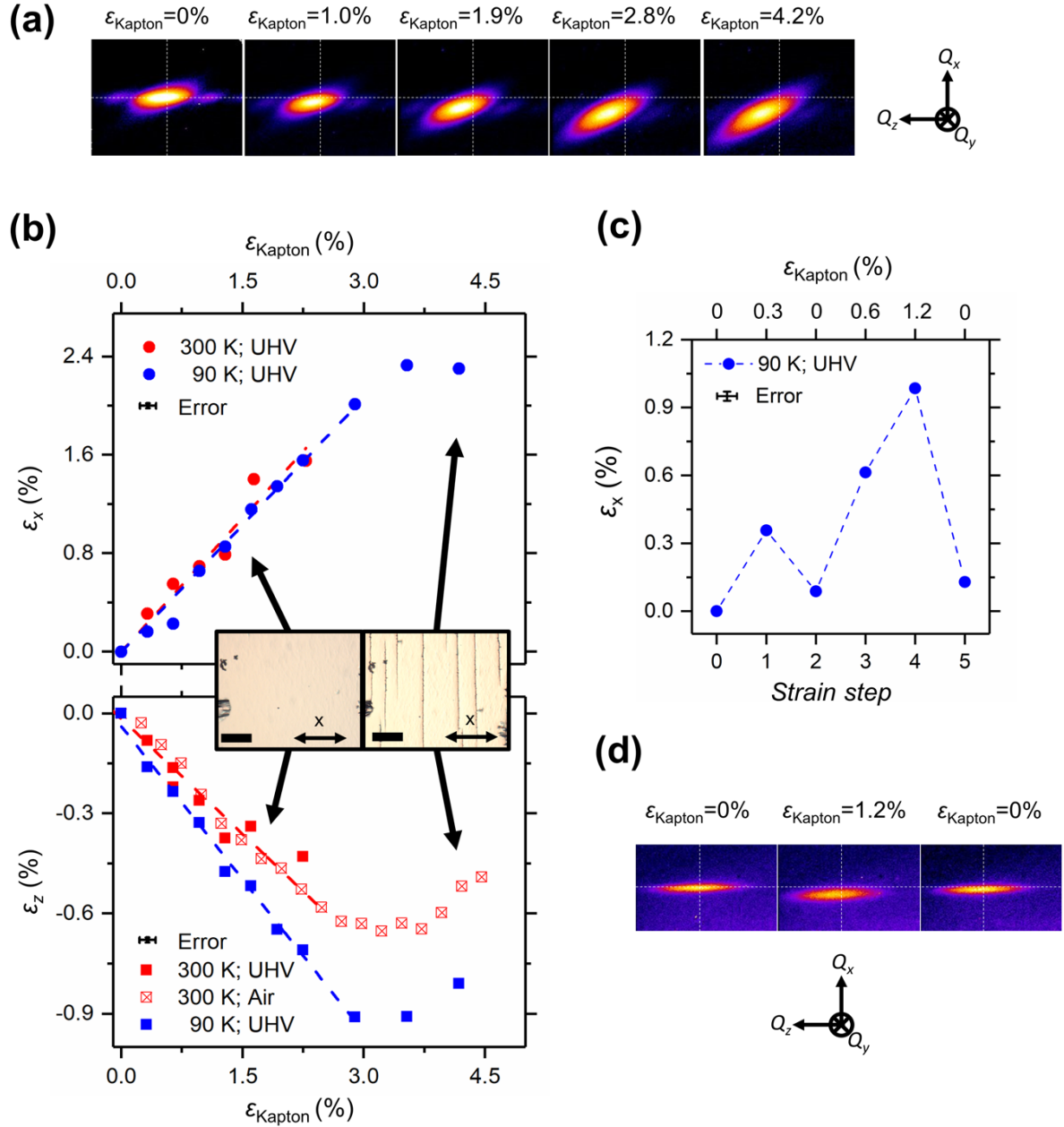

**Figure 2.** (a) Detector images showing the shift of the 015 reflection for increasing $\varepsilon_{Kapton}$. The intensity scale is quadratic. (b) Strain components of the $Bi_2Se_3$ film, $\varepsilon_x$ and $\varepsilon_z$, as a function of $\varepsilon_{Kapton}$ under different environments and at different temperatures. The strain in the $Bi_2Se_3$ films initially increases linearly before it starts to relax by defect formation. The inset presents optical micrographs demonstrating crack formation at large strains; scale bar 20 μm. (c) Measured $\varepsilon_x$ for cyclic straining and unstraining of the Kapton foil. (d) Detector images showing the reversible shift of the 015 reflection during cyclic straining and unstraining.

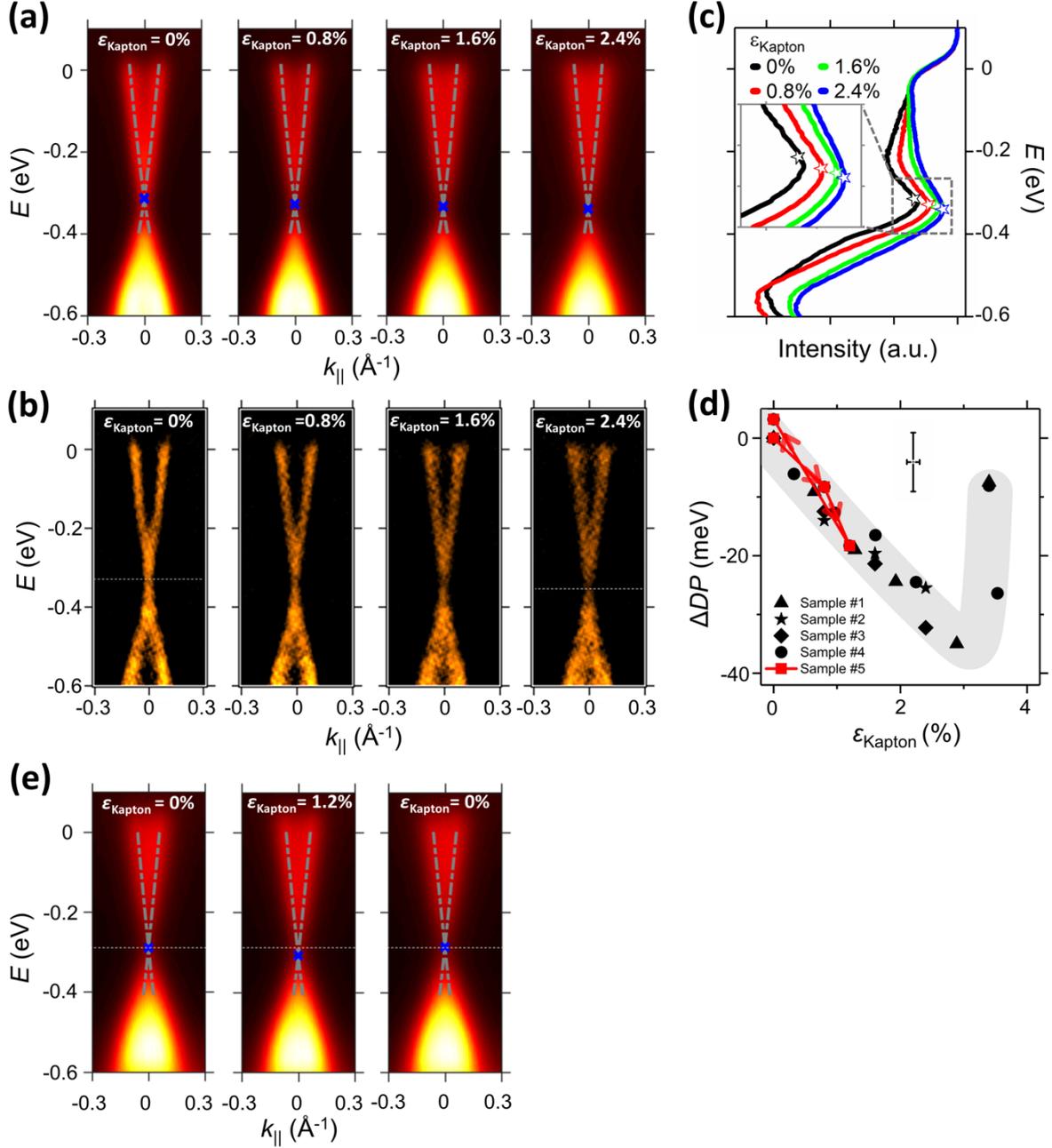

**Figure 3**. (a) ARPES maps along $\bar{\Gamma}$-$\bar{K}$ for a 10-QL $Bi_2Se_3$ film with increasing $\varepsilon_{Kapton}$ taken with 52 eV photons. Gray dashed curves show fitted dispersion relations of the topological surface states and the blue crosses mark the Dirac points. (b) Corresponding second-derivative ARPES maps. (c) Background corrected energy dispersion curves at $\bar{\Gamma}$ for various strain levels. Stars mark the Dirac points. (d) Shift of the Dirac point ($\Delta DP$) as a function of $\varepsilon_{Kapton}$ as measured during five independent tensile tests. The red dots are data from a straining/unstraining experiment and the grey line is a guidance to the eyes. (e) ARPES maps along $\bar{\Gamma}$-$\bar{K}$ during straining/unstraining, demonstrating the reversible shift of the Dirac point within the elastic limit.

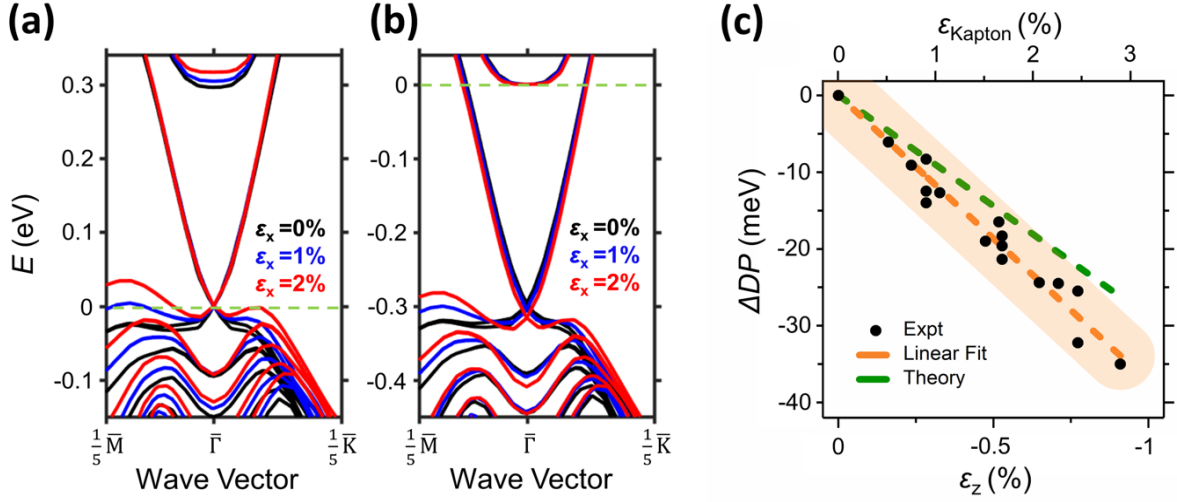

**Figure 4.** (a) Calculated band structures for a 6-QL Bi$_2$Se$_3$ film with the Dirac point set at $E = 0$ for $\varepsilon_x = 0$, 1, and 2%, emphasizing the upward shift of the conduction band minimum with increasing strain. (b) Same band structures plotted with the conduction band minimum set at $E = 0$ providing a direct comparison to the electron doped films in our experiment. (c) Experimental energy shifts (black dots) of the Dirac point relative to the Fermi level ($\Delta DP$) as a function of $\varepsilon_{Kapton}$ and $\varepsilon_z$. The orange dashed line is a linear fit to the data, and the orange shaded area indicates the error margins. Also shown is the theoretical prediction (green dashed line).

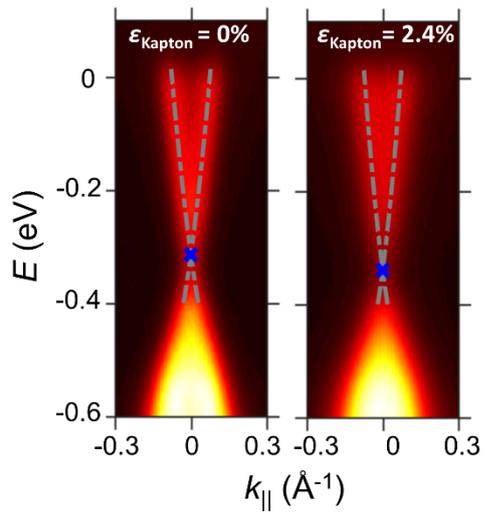
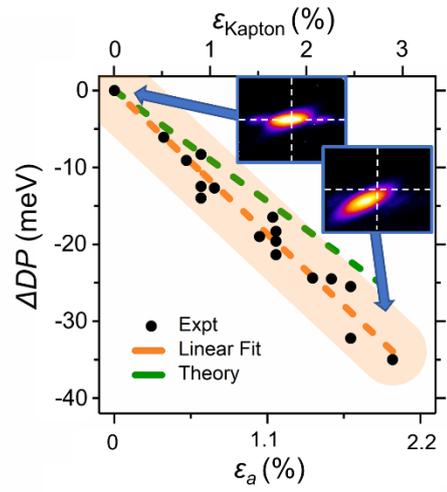

**Table of Contents Graphic**